\documentstyle[prb,twocolumn,aps]{revtex}

\tighten

\begin{document}
\title{Kondo effect and bistability in a double-quantum-dot}
\author{Pedro A. Orellana $^{1}$, G. A. Lara $^{2}$ and Enrique V. Anda $^{3}$}
\address{$^{1}${Departamento de F\'{\i}sica, Universidad Cat\'olica del Norte, }\\
Casilla 1280, Antofagasta, Chile.\\
$^{2}${Departamento de F\'{\i }sica, Universidad de Antofagasta,}\\
Casilla 170, Antofagasta, Chile.\\
$^{3}${Departamento de F\'{\i }sica, Pontificia Universidade Cat\'{o}lica do Rio de Janeiro, }\\
C.P.38071-970,Rio de Janeiro, RJ, Brazil.}
\maketitle

\begin{abstract}
We study theoretically the out-of-equilibrium transport properties of a
double quantum dot system in the Kondo regime. We model the system by means
of a two-impurity Anderson Hamiltonian. The transport properties are
characterized by Kondo effect properties, however, superimposed them, the
system possesses novel non-linear bistability behavior.
\end{abstract}

\smallskip

Recently experiments on quantum dots ($QDs$) at temperatures ($T$) below the
Kondo temperature ($T_{k}$) have shown that new physics emerges when their
transport properties are studied.\cite{goldhaber,cronenwett,lee} These
experiments confirm that many of the phenomena that characterize strongly
correlated metals and insulators, as it is the case of the Kondo effect, are
present in $QDs$. The advantage of studying the quantum-coherent many-body
Kondo state in $QDs$, in comparison with natural compounds, consists in the
possibility of continuous tuning the relevant parameters governing the
properties of this state, in an equilibrium and out-of-equilibrium
situation. In particular, the problem of electrons tunnelling through
double-quantum-dots ($DQDs$) in the Kondo regime has received much attention
in recent years.\cite{aguado,georges,busser,aono,izumida,ivanov} The $DQD$
is the simplest system where it is possible to study the competition between
the inter-dot antiferromagnetic spin-spin correlation and the dot-conduction
spin-spin correlation present in its ground state. The type of coupling
between the $QDs$ determines the character of the electronic states and the
transport properties of the artificial molecule. In the tunnelling regime,
the electronics states are extended across the entire system and form a
coherent state based on the bonding or anti-bonding levels of the $QDs$. In
this context, recently Aguado and Langreth \cite{aguado} studied the
out-of-equilibrium transport properties of a $DQD$ in the Kondo regime. They
found that for inter-dot-coupling greater than the level broadening, there
is a critical voltage above which the coherent configuration is unstable.
This instability is reflected as a drastic drop of the current leading to a
singular region of negative differential conductance. This behavior
resembles the $I$-$V$ characteristic of a double barrier structure in the
accumulation of charge regime. \cite{goldman} This system, due to
non-linearities introduced by the Coulomb interactions, has a bistable
behavior, characterized by two solutions for the current. In this work we
report the existence of a similar bistable behavior in an out-of-equilibrium
double quantum-dot in the Kondo regime.

We focus our study on a $DQD$ in the Kondo regime driven to an
out-of-equilibrium state by means of a dc voltage bias. In recent years the
very small width of fabricated semiconductor leads used to connect the $QDs$
permits to separate each lateral quantum confine energy level. In this case
the leads can be represented by a $1D$ tight-binding Hamiltonian connecting
the dots to two particle reservoirs characterized by Fermi levels $\mu _{L}$
and $\mu _{R}$ respectively. It has been shown that the physics associated
to a $DQD$ connected to two leads can be readily understood in terms of two
impurities Anderson Hamiltonian where the impurities are the $QDs$.\cite
{aguado,busser,aono} The Hamiltonian can be written as,

\smallskip

\begin{eqnarray}
&&H=\sum_{i\neq 0,1}\varepsilon _{i}n_{i\sigma }+t\!\!\!\sum_{\langle
i,j(\neq 0,1)\rangle ,\sigma }\ c_{i,\sigma }^{\dagger }c_{j,\sigma
}+\sum_{\alpha =0,1}\varepsilon _{\alpha }n_{\alpha ,\sigma }  \nonumber \\
&+&\frac{V_{L}}{\sqrt{2}}\sum_{\sigma }(c_{-1,\sigma }^{\dagger }c_{0,\sigma
}+c_{0,\sigma }^{\dagger }c_{-1,\sigma })+\frac{V_{R}}{\sqrt{2}}\sum_{\sigma
}(c_{1,\sigma }^{\dagger }c_{2,\sigma }+c_{2,\sigma }^{\dagger }c_{1,\sigma
})  \nonumber \\
&+&\frac{t_{c}}{2}\sum_{\sigma }(c_{0,\sigma }^{\dagger }c_{1,\sigma
}+c_{1,\sigma }^{\dagger }c_{0,\sigma })+U\sum\limits_{\alpha =0,1}n_{\alpha
_{\uparrow }}n_{\alpha _{\downarrow }},
\end{eqnarray}
\noindent where the dots have been localized at sites $0$ and $1$ of the
lattice. The operator $c_{i\sigma }^{\dagger }$ creates an electron in the
site $i$ with spin $\sigma $, $\varepsilon _{i}$ is the site energy, $t$ is
the hopping in the leads, $V_{L(R)}$ is the hopping between the left (right)
leads and the left (right) $QD$, $t_{c}$ is the inter-dot coupling
tunnelling and $U$ is the on-site Coulomb energy. In typical experiments,
the intra-dot Coulomb repulsion $U$ is large compared with $kT$. In this
case, taking the limit $U\rightarrow \infty ,$ it is possible to obtain an
important simplification of the model because the double occupancy at the
dots is eliminated from the Hilbert space.

The Hamiltonian may be written in terms of auxiliary slave boson operators
plus constraints.\cite{barnes,coleman,read}The annihilation operator of an
electron at dot $\alpha $ is decomposed as $c_{\alpha \sigma }=b_{\alpha
}^{\dagger }f_{\alpha \sigma }$, where a slave boson operator $b_{\alpha
}^{\dagger }$ creates an empty state and a fermion operator $f_{\alpha
\sigma }$ annihilates a single occupied state with spin $\sigma $. The sites 
$\alpha =0,1$ can only be in one of the states $b_{\alpha }^{\dagger }\left|
0\right\rangle $ or $f_{\alpha \sigma }^{\dagger }$ $\left| 0\right\rangle $%
. The slaves boson operator acts preventing double occupancy of the site.
When an electron creation operator acts on an occupied site, the boson part
annihilates the state; $c_{\alpha ,\sigma }^{\dagger }$ $f_{\alpha \sigma
}^{\dagger }$ $\left| 0\right\rangle =f_{\alpha ,\sigma }^{\dagger
}b_{\alpha }$ $f_{\alpha \sigma }^{\dagger }\left| 0\right\rangle =0.$

The exclusion of double occupancy at site $\alpha $, imposes the condition, $%
Q_{\alpha }\equiv \sum\nolimits_{\sigma }f_{\alpha \sigma }^{\dagger
}f_{\alpha \sigma }+b_{\alpha }^{\dagger }b_{\alpha }=1,$ for the number of
bosons and fermions at that site. \cite{read} This constraint can be taken
into account by adding a term, $\sum\nolimits_{\alpha }\lambda _{\alpha
}(Q_{\alpha }-1)$, to the Hamiltonian (Eq.1) with the Lagrange multipliers $%
\lambda _{\alpha },$that are calculated imposing the constraint condition.
In order to solve the Hamiltonian we adopt the mean field approximation
(MFA) where the Bose operators $b_{\alpha }^{\dagger }$ and $b_{\alpha }$
are replaced by the expectation values, $\left\langle b_{\alpha
}\right\rangle =\widetilde{b}_{\alpha }\sqrt{2}=\left\langle b_{\alpha
}^{\dagger }\right\rangle =\widetilde{b}_{\alpha }^{\dagger }\sqrt{2}$,\cite
{read} neglecting its fluctuations.  

\smallskip Adopting the slave-bosons $MFA$, the tight-binding two-impurity
Anderson Hamiltonian describing a double quantum-dot connected to leads can
be written as, 
\begin{eqnarray}
H &=&\sum_{i}\varepsilon _{i}n_{i\sigma }+t\sum_{<ij\neq 0,1>\sigma
}c_{i\sigma }^{\dagger }c_{j\sigma }+\sum\limits_{\alpha =0,1,\sigma }%
\widetilde{\varepsilon }_{\alpha }n_{\alpha \sigma }  \nonumber \\
&+&\widetilde{V}_{L}\sum_{\sigma }(c_{-1\sigma }^{\dagger }f_{0\sigma
}+f_{-1\sigma }^{\dagger }c_{0\sigma })+\widetilde{V}_{R}\sum_{\sigma
}(f_{1\sigma }^{\dagger }c_{2\sigma }+f_{1\sigma }^{\dagger }c_{2\sigma }) 
\nonumber \\
&+&\widetilde{t}_{c}\sum_{\sigma }(f_{0\sigma }^{\dagger }f_{1\sigma
}+f_{1\sigma }^{\dagger }f_{0\sigma })+\sum\nolimits_{\alpha }\lambda
_{\alpha }(\widetilde{b}_{\alpha }^{\dagger }\widetilde{b}_{\alpha }-1).
\end{eqnarray}
\mathstrut

\smallskip \noindent where $\widetilde{\varepsilon }_{0}=\varepsilon
_{0}+\lambda _{0}$, $\widetilde{\varepsilon }_{1}=\varepsilon _{1}+\lambda
_{1}$, $\widetilde{V}_{L}=V_{L}\widetilde{b}_{0}$, $\widetilde{V}_{R}=V_{R}%
\widetilde{b}_{1}$, $\widetilde{t}_{c}=t_{c}\widetilde{b}_{0}\widetilde{b}%
_{1}$.

On a tight-binding basis, the stationary state of energy $\varepsilon _{k}$
results to be 
\begin{equation}
\left| \psi _{k\sigma }\right\rangle =\sum_{i}a_{i\sigma }^{k}\left| \phi
_{i\sigma }\right\rangle
\end{equation}
\noindent where $\left| \phi _{i\sigma }\right\rangle $ is a Wannier state
localized at site $i$ of spin $\sigma $ and the coefficients $a_{i\sigma
}^{k}$ obey the non-linear difference equations,

\begin{eqnarray}
\varepsilon _{k}a_{j,\sigma }^{k} &=&\varepsilon _{j}a_{j,\sigma
}^{k}+t(a_{j-1,\sigma }^{k}+a_{j+1,\sigma }^{k})\;\;\;\;(j\neq -1,0,1,2), 
\nonumber \\
\varepsilon _{k}a_{-1(2),\sigma }^{k} &=&\varepsilon _{-1(2)}a_{-1(2),\sigma
}^{k}+\widetilde{V}_{L(R)}a_{0(1),\sigma }^{k}+ta_{-2(3),\sigma }^{k}, 
\nonumber \\
\varepsilon _{k}a_{0(1),\sigma }^{k} &=&\widetilde{\varepsilon }%
_{0(1)}a_{0(1),\sigma }^{k}+\widetilde{V}_{L(R)}a_{-1(2),\sigma }^{k}+%
\widetilde{t}_{c}a_{1(0),\sigma }^{k},
\end{eqnarray}

\noindent where $a_{j,\sigma }^{k}$ is the amplitude of probability to find
the electron in the site $j$ in the state $k$ with spin $\sigma .$ The mean
field parameters ($\widetilde{b}_{0}$, $\widetilde{b}_{1}$, $\lambda _{0}$, $%
\lambda _{1}$) are determined by minimizing the free energy of the systems.
Taking the expectation value of the Hamiltonian, differentiating and using
the Hellman-Feymann theorem,\cite{read}it is possible to find that they
satisfy the set of four equations:

\begin{eqnarray}
\widetilde{b}_{0(1)}^{2}+\frac{1}{2}\sum\limits_{k,\sigma }\left|
a_{0(1),\sigma }^{k}\right| ^{2} &=&\frac{1}{2},  \nonumber \\
\widetilde{V}_{L(R)}\sum\limits_{k,\sigma }{%
\mathop{\rm Re}%
}(a_{-1(2),\sigma }^{k*}a_{0(1),\sigma }^{k}) &+&  \nonumber \\
\widetilde{t}_{c}\sum\limits_{k,\sigma }{%
\mathop{\rm Re}%
}(a_{1(0),\sigma }^{k*}a_{0(1),\sigma }^{k})+\lambda _{0(1)}\widetilde{b}%
_{0(1)}^{2} &=&0.
\end{eqnarray}

\noindent The sum over the wave vector $k$ cover all the occupied states.
The resulting equations are nonlinear because of the renormalization of the
localized levels in the dots, the inter-dots coupling tunneling and the
coupling tunneling between the QDs and the leads. Due to the presence of
these non-linear terms, the set of equations (4) and (5) have to be solved
self-consistently.

In order to study the solutions of equations (4) and (5) we assume that the
electrons are described by a plane wave incident from the far left with an
amplitude $I$ and a reflection amplitude $R$ and at the far right by a
transmission amplitude $T$. Taking this to be the solution we can write, 
\begin{eqnarray}
a_{j}^{k} &=&Ie^{ikr_{j}}+Re^{-ikr_{j}},j<-2  \nonumber \\
a_{j}^{k} &=&Te^{ikr_{j}},j>3
\end{eqnarray}
\noindent where $r_{j}$ is the position of site $j$. The solution of
equations (4) can be obtained through an adequate iteration of it from right
to left. For a given transmitted amplitude, the associated reflected and
incident amplitudes may be determined by matching the iterated function to
the proper plane wave at the far left. The transmission probability $\left|
T/I\right| ^{2}$, obtained from the iterative procedure, multiplied by the
wave vector $k$, gives us the contribution of this wave vector to the
current. With the purpose of solving equations (4) an (5) we define a second
pseudo-time-like iteration in the following way. Initially Eqs. (4) are
solved, for a particular applied potential, $V,$ ignoring the non-linear
term and for energies up to the maximum Fermi energy. The coefficients thus
obtained correspond to a solution for non-interacting electrons. They are
used to construct the non-linear term for the next solution. The procedure
is repeated up to the moment in which convergence is reached. This solution
is taken as the starting point for the next cycle corresponding to another
value of $V.$

Once the amplitudes $a_{j,\sigma}^{k}$ are known, the current is numerically
obtained from,

\begin{equation}
J=\frac{2e}{\hbar }\tilde{V}_{L}\sum\limits_{k,\sigma }{%
\mathop{\rm Im}%
}\{a_{-1,\sigma }^{k*}a_{0,\sigma }^{k}\}.
\end{equation}

We study a model which consists of two leads connected to quantum dots with $%
\mu _{L}=-V/2$ and $\mu _{R}=V/2$, $t_{\text{ }}=30\;\Gamma $, $%
V_{L}=V_{R}=V_{0}=5.48\;\Gamma $, $\varepsilon _{0}=\varepsilon
_{1}=-3.5\;\Gamma $ (Kondo regime with $T_{K}\approx 10^{-3}\;\Gamma $ with $%
\Gamma =\pi V_{0}^{2}\rho (0)$). The normalization of the wave function is
taken so that each site of the leads can be populated by a maximum of two
electrons.

Fig 1 shows the $I$-$V$ characteristics. The system has a bistable behavior,
with two solutions for the current, within a range of values for the
external potential that increases with $t_{c}$ . These solution are obtained
as the voltage is increased (solid line), or decreased (dashed line). When
it is increased above a critical value $V_{c_{\uparrow }}$, the coherence
between the dots is lost and the current drops.\cite{aguado} On the other
hand, when it is reduced from above $V_{c_{\uparrow }}$, the Kondo
resonances in each dot are pinned at their own chemical potential and it is
necessary to reduce the voltage by an additional amount $\Delta V$, below $%
V_{c_{\uparrow }}$, to restore the coherence between the dots. The width of
the bistability $\Delta V$ is proportional to $t_{c}$ and it is about $%
\Delta V\sim 1\mu eV($ $t_{c}=1.3\Gamma )$, that is within the resolution
limits of the experiments. \cite{goldhaber}.

It is well known that the slave-boson $MFA$ describes a system that, for a
region of the parameters space or for $T>T_{k}$, possess a non-physical
solution that disconnects the magnetic impurity from the rest of the system.
For the problem we are studying of a $DQD$ this solution coexist with the
physical one \cite{Jones}. This non-physical behavior is not associated to
the bistability we are describing. To make this point clear in Fig.1 we
represent the spurious solution (dot line), as a function of the external
potential with $I=0$ for $V$ $\leq V_{c}^{*}$, $I\neq 0,V$ $>V_{c}^{*}$.

For $t_{c}>\Gamma $, the Kondo resonances are split into a bonding and
antibonding peaks. They are located at, $\widetilde{\varepsilon }_{\pm }={(%
\widetilde{\varepsilon }_{0}+\widetilde{\varepsilon }_{1})\pm \sqrt{(%
\widetilde{\varepsilon }_{0}-\widetilde{\varepsilon }_{1})^{2}+4\widetilde{t}%
_{c}^{2}}\}/2}$. In fig 2 we plot $\widetilde{\varepsilon }_{\pm }$ and $%
\widetilde{\varepsilon }_{0(1)}$ as a function of $V$ for $t_{c}=1.3\Gamma $%
. For $V=0$, the bonding and antibonding levels satisfies $\widetilde{%
\varepsilon }_{\pm }$ $=$ $\pm \widetilde{t}_{c}$and the site energies $%
\widetilde{\varepsilon }_{0(1)}=0.$ When the bias $V$ is increased, $%
\widetilde{\varepsilon }_{\pm }$ are kept almost constant and $\widetilde{%
\varepsilon }_{0\text{ }}$ and $\widetilde{\varepsilon }_{1\text{ }}$
decreases and increases respectively. For higher values of $V$ (always $%
V<V_{c_{\uparrow }}$) , $\widetilde{\varepsilon }_{+\text{ }}$and $%
\widetilde{\varepsilon }_{-\text{ }}$approach each other, $\widetilde{%
\varepsilon }_{0\text{ }}$goes up and $\widetilde{\varepsilon }_{1\text{ }}$%
down, until they almost coincide again at $V=V_{c_{\uparrow }}$. For a
voltage close to and above $V_{c_{\uparrow }}$, $\widetilde{\varepsilon }%
_{\pm }$ and $\widetilde{\varepsilon }_{0(1)}$ converge towards their own
chemical potential $\mu _{L(R)}$. When the external potential is decreased
from above $V_{c_{\uparrow }}$, the $\widetilde{\varepsilon }_{\pm }$ are
kept near their own chemical potential until a critical voltage $%
V_{c_{\downarrow }}$is reached, where they recover abruptly the values
obtained when the potential was increased from zero. Decreasing the voltage
in the interval $V_{c_{\downarrow }}<V<V_{c_{\uparrow }}$, the separation
between the $\widetilde{\varepsilon }_{\pm }$ is less than $V$ and reduces
with $V$ until $V=V_{c_{\downarrow }}$. Below $V_{c_{\downarrow }}$, the
system returns to a coherent state and the $\widetilde{\varepsilon }_{\pm }$
go again to the neighborhood of $\pm \widetilde{t}_{c}$.

\medskip Fig 3 shows the left and right $DOS$ of the $QDs\ $for $%
t_{c}=1.3\Gamma $ and a voltage inside the bistability region ($V=2.15T_{k})$%
. The figure shows a coherent solution with the dots strongly coupled (high
current; solid line), with an equally shared split Kondo peak and another
one with the dots weakly coupled (low current; dashed line), each having a
DOS with a Kondo peak pinned at their corresponding Fermi energies.

The nonlinear behavior described above has similarities with the well-known
bistability present in $3D$ resonant tunneling double barrier systems and
multistabilities in doped superlattices. \cite
{goldman,anda,orellana,prengel,kastrup,aguado2}. In $3D$ tunneling double
barrier systems, the Coulomb interaction pins the renormalized energy level
at the well when the applied voltage $V$ is augmented maintaining the system
at resonance. However, when $V$ goes above a critical value the abrupt
leakage of the charge accumulated in the well takes the system out of
resonance. For doped superlattices, at large carrier densities, the internal
potential profile produced by the applied voltage breaks up into domains
giving rise to a multistable behavior in the current\cite
{prengel,kastrup,aguado2}. In both systems, the non-linear deformation of
the potential profile due to the electron-electron interaction is
responsible for the phenomenon. Although in our case the Kondo effect is a
strong many body effect of a different nature in comparison with the
previous systems, the bistability is as well a consequence of the non-linear
potential profile deformations that the electrons suffer when the external
potential is modified. Increasing the voltage (always $V$ $<V_{c_{\uparrow }}
$ ), the interaction maintains the site energy of the right dot above the
left dot level as shown in Fig $2$. For these values of $V$, the electron
wave function delocalizes and spreads on the double-quantum dot. In contrast
just above $V_{c_{\uparrow }},$ this configuration becomes unstable, the
coherence between the dots is lost, their interaction reduces abruptly as
the external potential is augmented and the wave function localizes at each
quantum dot. The tunneling coupling between the dots ($\widetilde{t}_{c}$)
reduces abruptly at  $V_{c_{\uparrow }}$  and consequently the current
circulating along the system (Fig $4$). For the $DQD$ as it was as well for
the double barrier heterostructure and the superlattices, due to the
nonlinearities produced by the Coulomb interaction, this transition occurs
at different values of $V$, depending wether the potential is increased or
decreased, giving rise to the bistability structure present in the $I$-$V$
characteristics.

In conclusion, we have reported that a double quantum-dot could have a
bistable behavior in the Kondo regime at zero temperature. This evidence was
obtained using a mean-field approach that neglects fluctuations, appropriate
for large spins as it is the first term of an $1/N$ expansion, where $N$ is
the spin dimension. Some experiments have shown\cite{Sasaki} that it is
possible to design dots with a configuration that, due to the exchange
interaction, the parallel spin coupling following Hund's rule gives rise to
a total dot spin $S_{t}>1/2$. For this cases our results are more reliable
as fluctuations are less important. Although we believe that our results are
robust against them, a study that goes beyond mean field is necessary to
confirm our conclusions. As mentioned above the slave boson $MFA$ have a
non-physical solution and it is possible to speculate that the negative
differential resistance and the bistable behavior of the current could be
associated to it. However, we were able to show that the phenomena here
studied are not associated to this artifact of the mean field approximation
as they correspond to different solutions of the problem.

Although bistabilities are a well-known phenomenon in resonant tunneling, to
the best of our knowledge neither its existence has been proposed nor has it
been measured in a $1D$ Kondo system. This phenomena is expected to occur
due to the high nonlinearity of the problem. Other possible manifestations
arising from nonlinearity in $DQD$ are multistabilities, self-sustained
oscillations and chaos phenomena, that are currently been investigated.

Work supported by grants Milenio ICM P99-135-F, FONDECYT\ 1990443,
C\'{a}tedra Presidencial en Ciencias, Fundaci\'{o}n Antorchas/Vitae/Andes ,
FINEP and CNPq.

\begin{figure}[tbp]
\caption{$I-V$ curve for a) $t_{c}=1.0\Gamma$, b)$t_{c}=1.1\Gamma$, c)$%
t_{c}=1.2\Gamma$ and d) $tc=1.3\Gamma$}
\end{figure}

\begin{figure}[tbp]
\caption{Level energies, a) $\epsilon_{\pm}$ vs $V$ and b) $\epsilon_{0(1)}$
vs $V$ for $t_c=1.3\Gamma$}
\end{figure}

\begin{figure}[tbp]
\caption{Local density of states for left and righ dot for $t_c=1.3$ and $%
V=2.15T_k$}
\end{figure}

\begin{figure}[tbp]
\caption{a) Renormalized level broadening in the left and the right dots as
function of the voltage, b) the renormalized interdot coupling tunneling
versus voltage for $t_c=1.3 \Gamma$}
\end{figure}

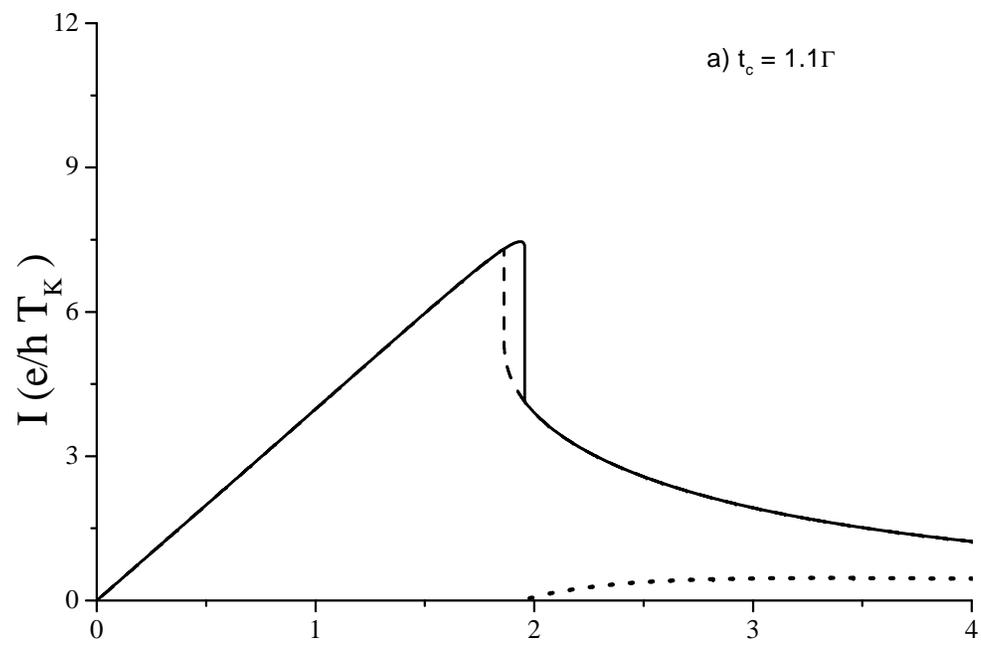
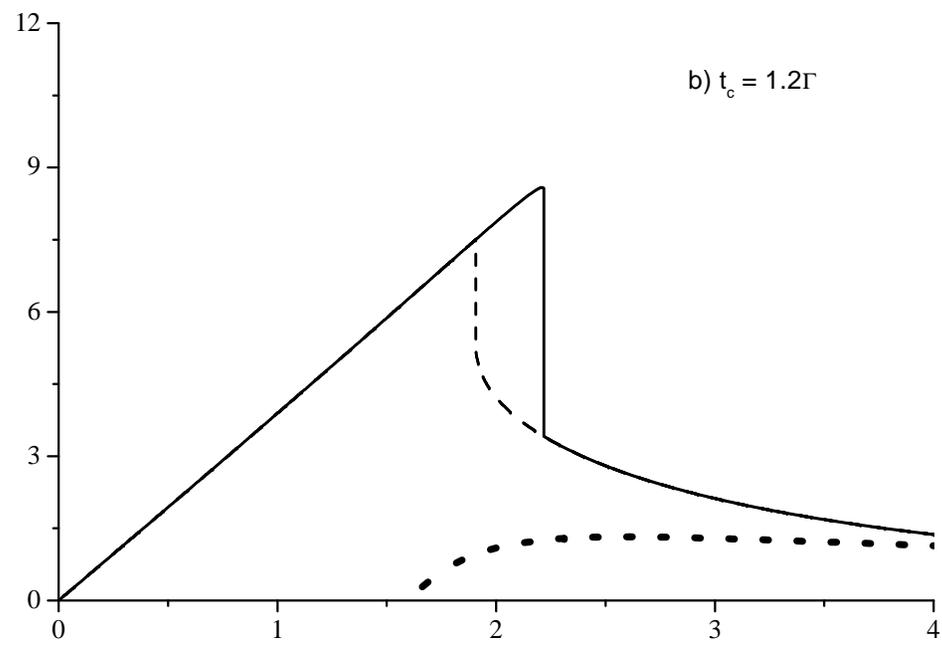
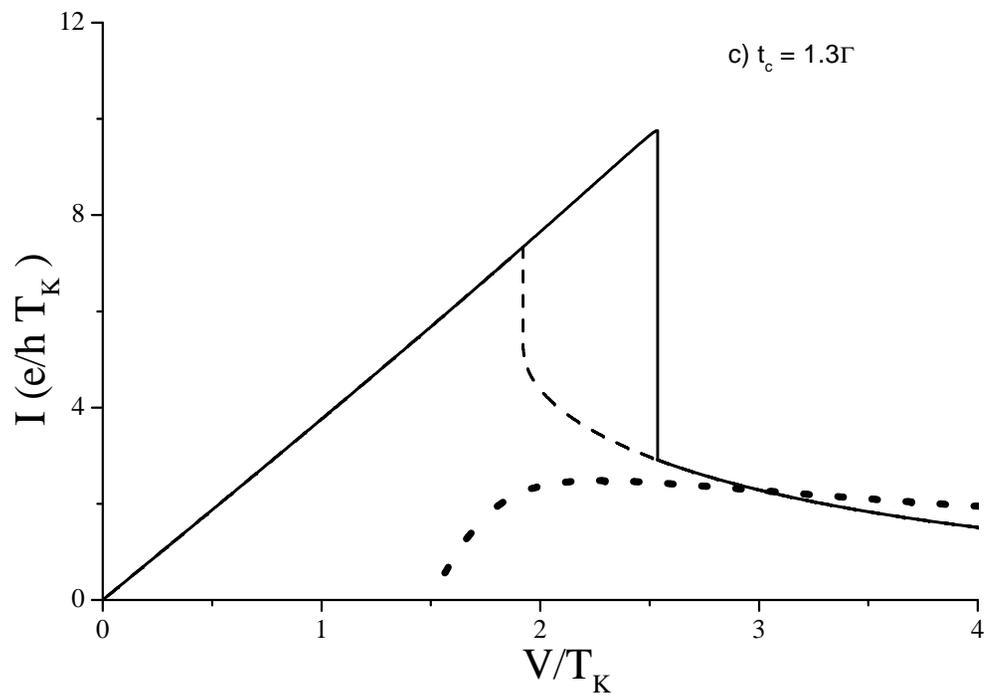
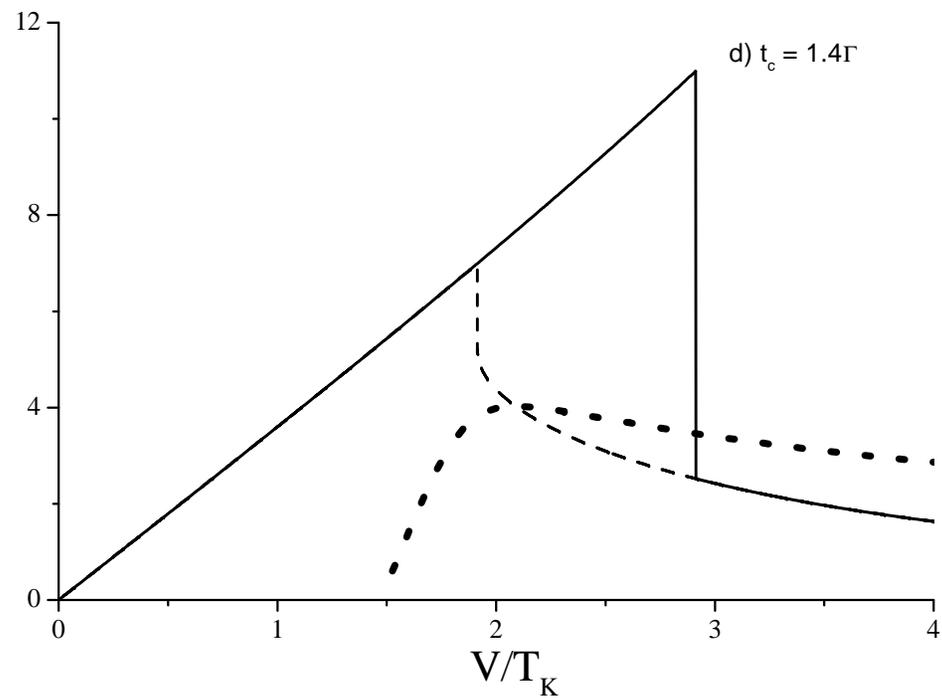
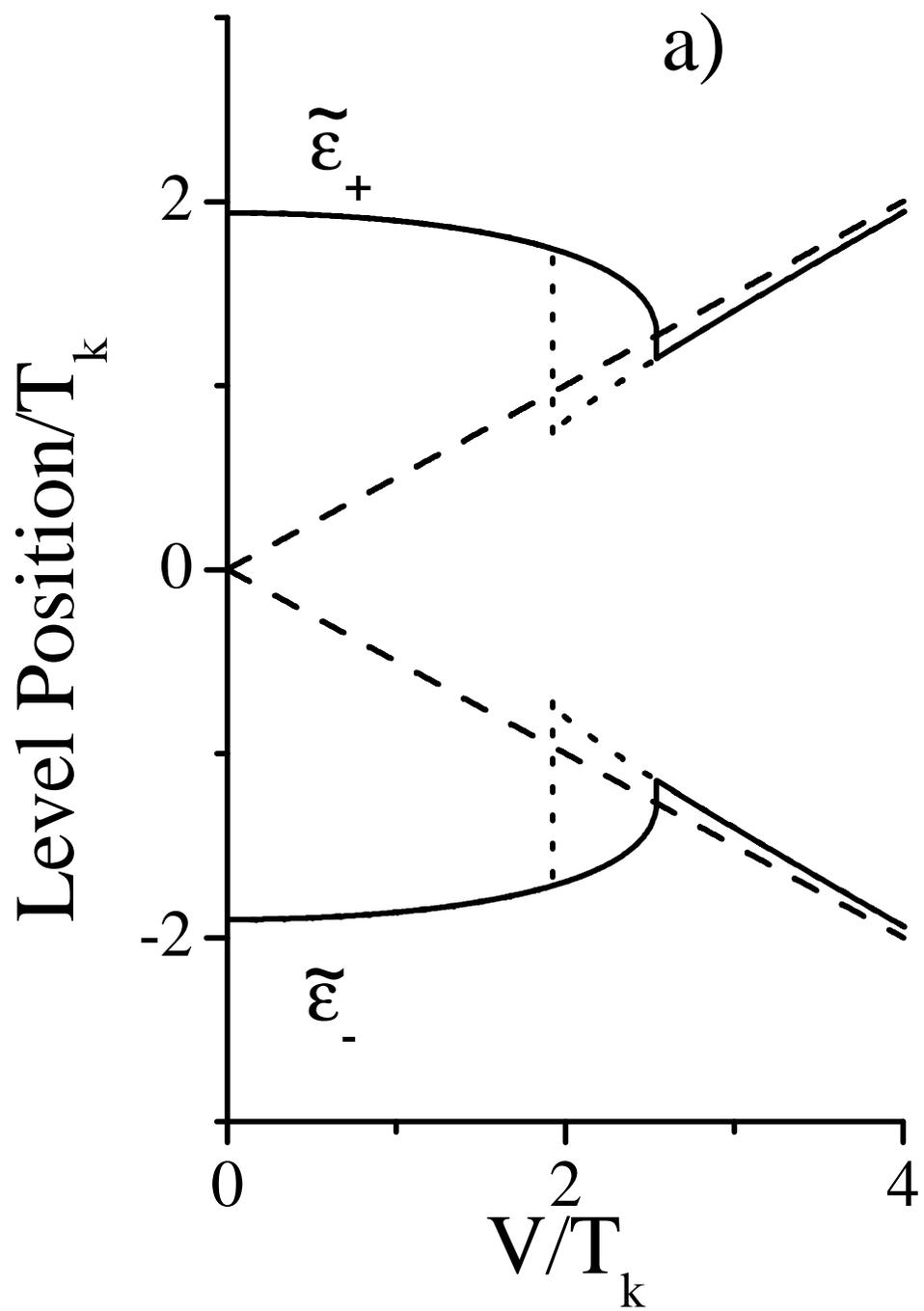
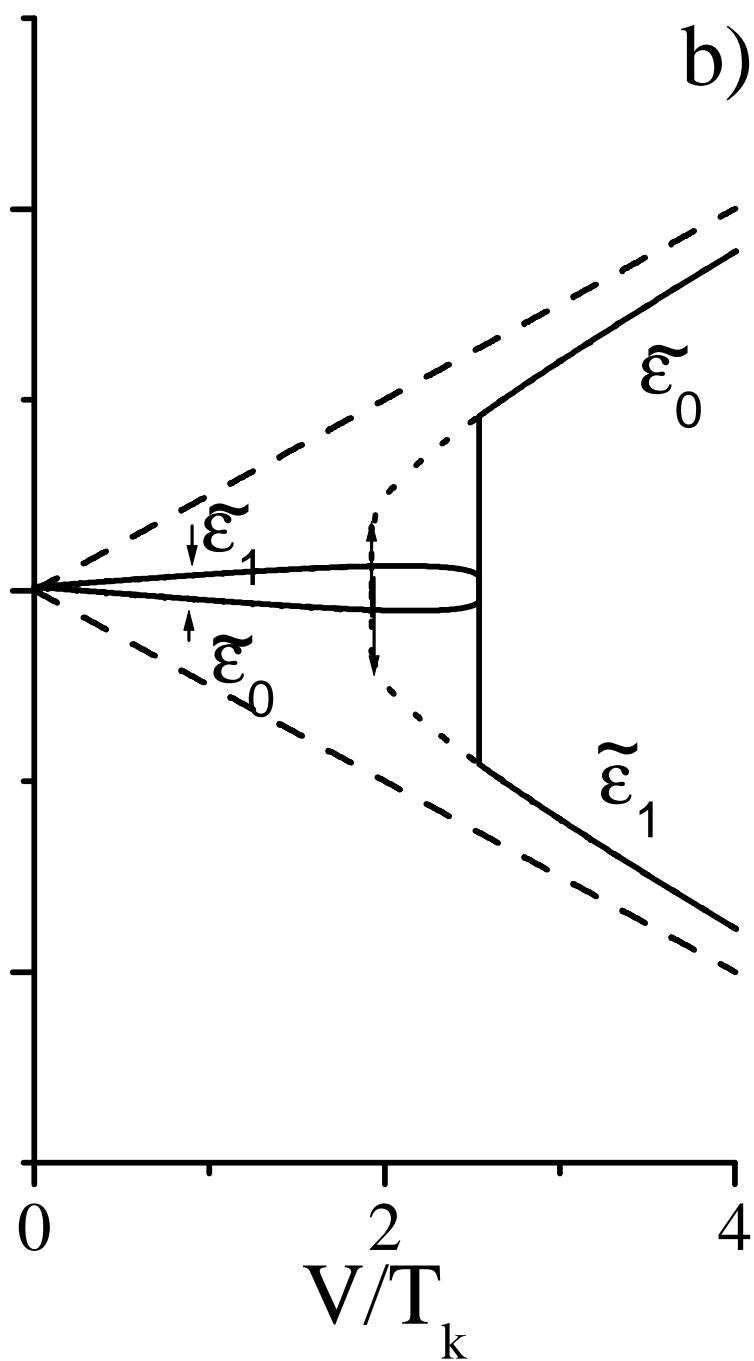
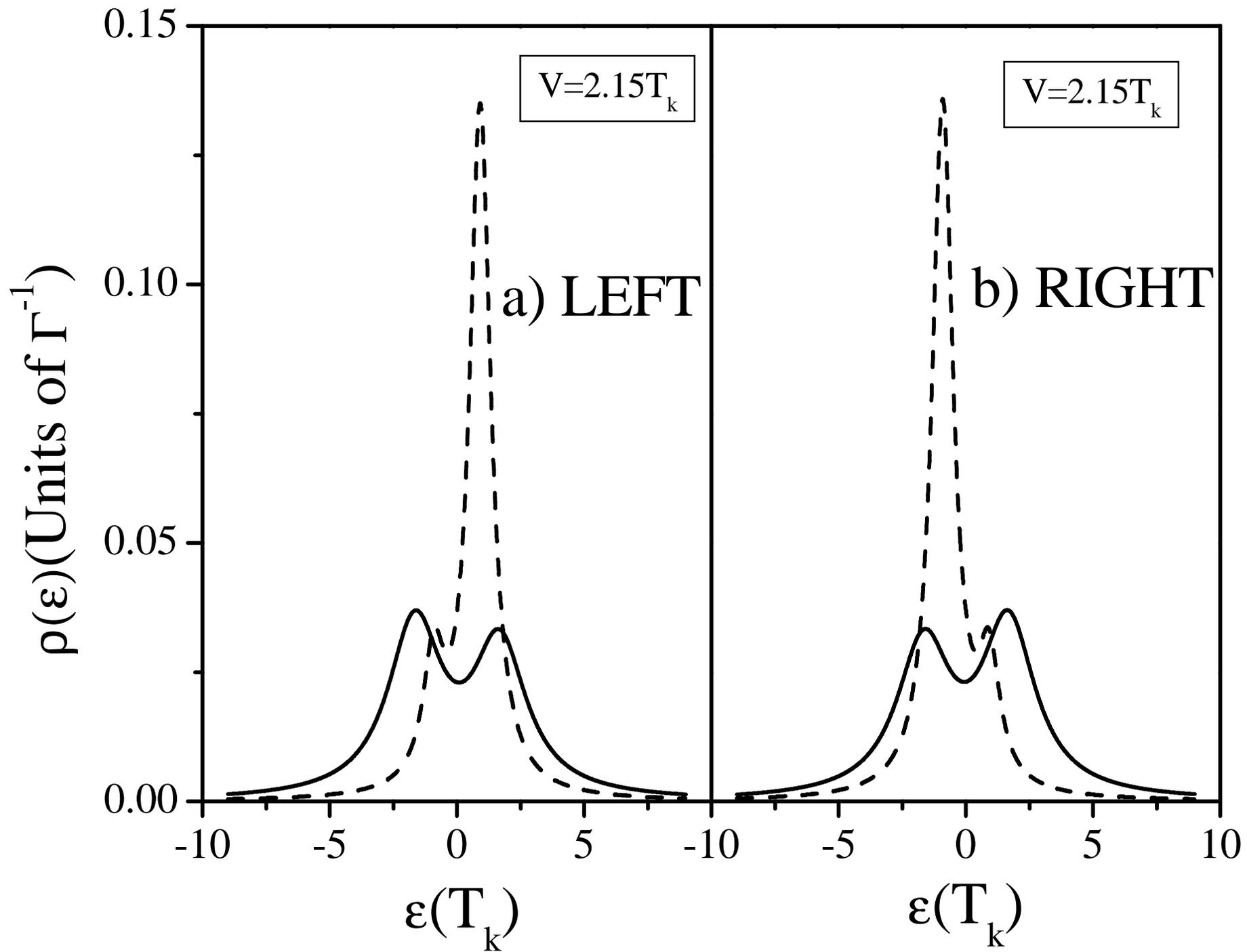
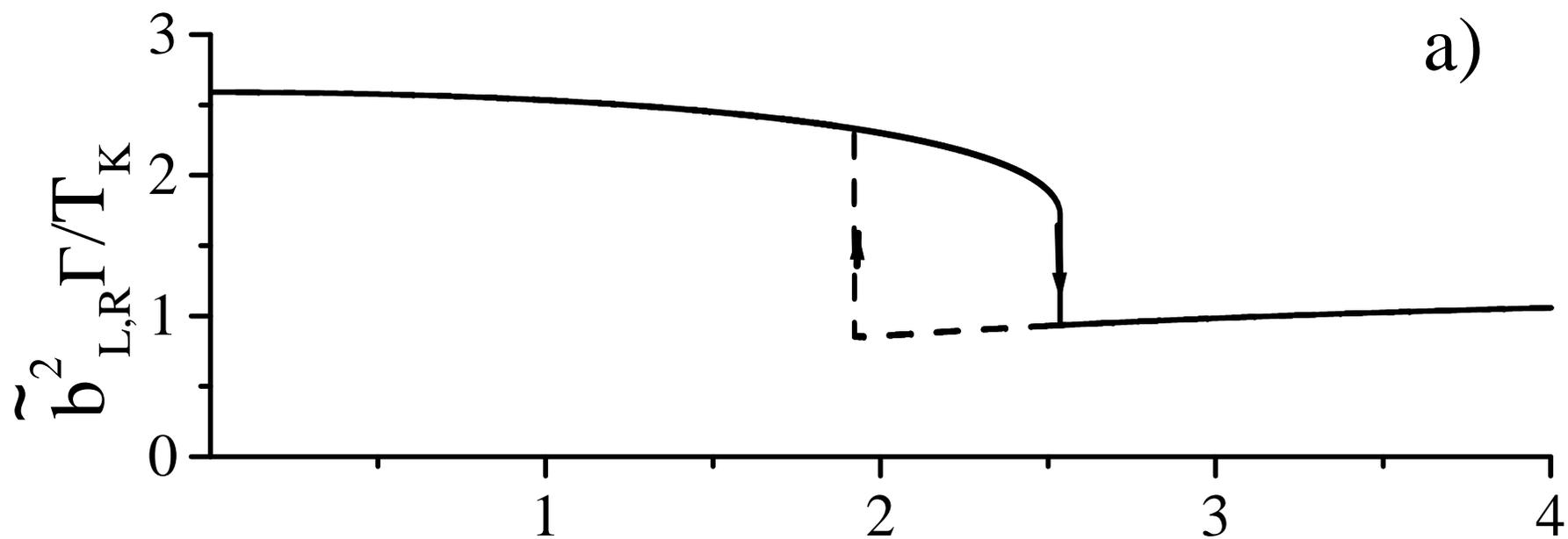
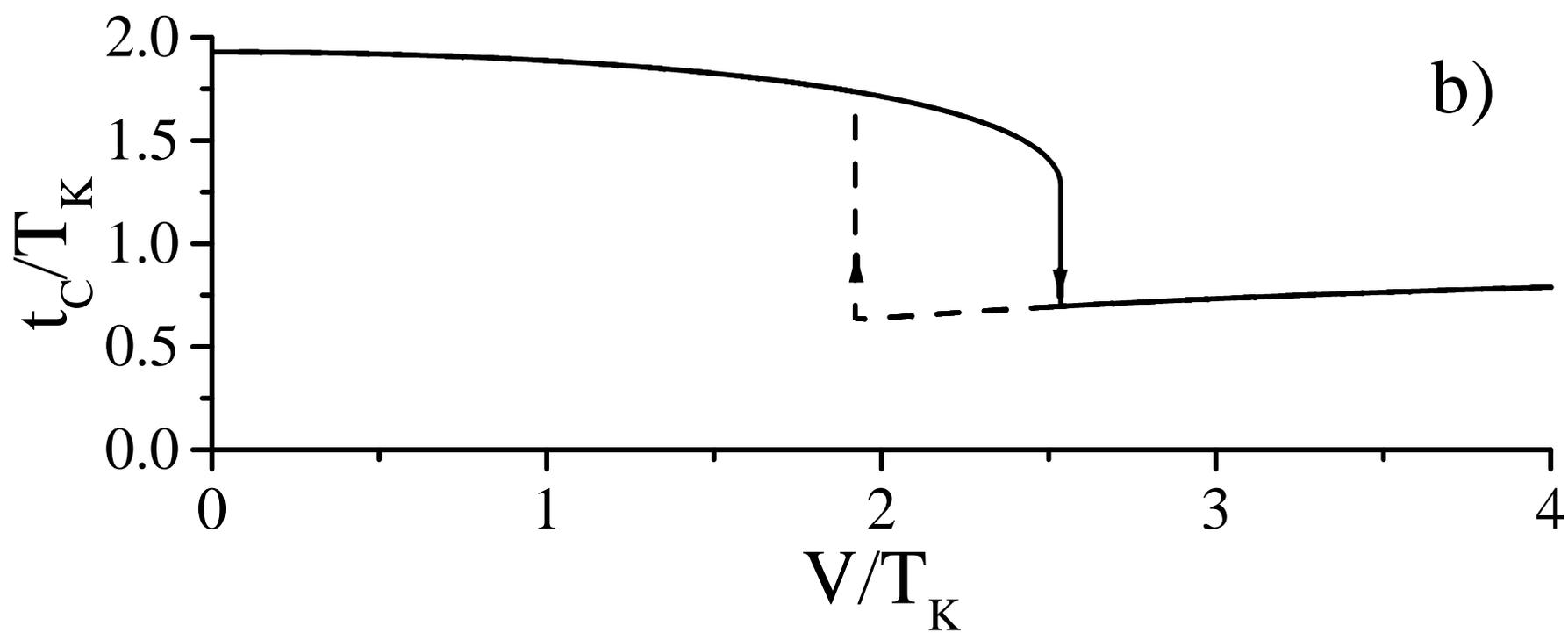
\end{document}